\documentclass[aps,pra,twocolumn,superscriptaddress,altaffilletter,secnumarabic,nobalancelastpage,amsmath,amssymb,nofootinbib,longbibliography,floatfix]{revtex4-2}

\usepackage[]{mdframed}
\usepackage{parskip}
\usepackage[shortlabels]{enumitem}
\usepackage{setspace}
\usepackage{framed}
\usepackage[version=3]{mhchem} 
\usepackage[utf8]{inputenc}   
\usepackage{CJK}
\usepackage{enumitem}
\usepackage{graphics}   
\usepackage{empheq} 
\usepackage[caption=false]{subfig}
\usepackage{comment}
\usepackage{tikz}
\usepackage{graphicx,float}      
\usepackage{url}           
\usepackage{bm}     
\usepackage{subfig}
\usepackage{caption}
\usepackage{mwe}
\usepackage{xcolor}

\usepackage[T1]{fontenc}
\usepackage{amssymb}
\usepackage{amsmath}
\usepackage{amsthm}
\usepackage{amsfonts}
\usepackage{lipsum}

\usepackage{float}
\usepackage{enumitem}
\usepackage[usestackEOL]{stackengine}

\begin{document}
\linespread{0.75}
\title{Magnetised Dense Nuclear Matter in Neutron Stars: A Relativistic Mean-Field Study of the Equation of State}
\author	   {Mojdeh Banafsheh}
\email          {mbanafsheh@ucmerced.edu}
\affiliation    {Department of Physics, university of California, Merced, Merced, CA 95343, USA}

\begin{abstract}
We study the static response of dense neutron-star matter to a prescribed magnetic-field strength within a deliberately minimal and internally consistent relativistic mean-field (RMF) baseline. The model includes neutrons, protons, electrons, and muons in beta equilibrium and charge neutrality, with a uniform external magnetic field incorporated through Landau quantisation of the charged species. The equation of state is evaluated self-consistently at zero temperature, while moderate finite-temperature effects are estimated through leading-order degenerate Sommerfeld corrections. Magnetic pressure anisotropy is included through the Maxwell contribution in the no-magnetisation approximation.

The purpose of this baseline is to isolate the hierarchy of magnetic-field effects before introducing additional microphysics or dynamical magnetic-field evolution. We compare the linear QHD-I parameterisation, used as a stiff benchmark, with the nonlinear GM1 model as a more realistic reference. The results show that canonical magnetar-scale fields produce negligible changes in the bulk core equation of state, while visible Landau-quantisation and pressure-anisotropy effects emerge only as the field approaches the strongly quantising regime. The comparison between QHD-I and GM1 further shows that nuclear-model dependence dominates over static magnetic-field corrections up to the field strengths explored here.

As an exploratory diagnostic, we use the isotropised equation of state to estimate the sensitivity of ordinary TOV mass--radius sequences to the prescribed magnetic field. This should not be interpreted as a fully anisotropic magnetised-star calculation. Anomalous magnetic moments, hyperons, quark degrees of freedom, and dynamical magnetic-field evolution are left for future extensions.
\end{abstract}
\maketitle

\section{Introduction}

Neutron stars provide a unique environment for studying
dense nuclear matter under extreme conditions of density,
temperature, and magnetic field. In magnetars, surface
magnetic fields can reach \(10^{14}\)--\(10^{15}\,\mathrm{G}\),
while even larger values may occur in the stellar interior\cite{turolla2015magnetars, duncan1992formation}.
Such fields affect the thermodynamics of charged particles through
Landau quantisation, modify the equation of state (EoS),
and introduce pressure anisotropy in magnetised matter.
These effects are relevant to neutron-star structure, thermal
evolution, and observable signatures such as X-ray emission
and cooling behaviour. Warm magnetised stellar matter has
also been investigated in protoneutron-star contexts, where
thermal and neutrino effects can become important~\cite{rabhi2011warm}.

Relativistic mean-field (RMF) models provide a widely
used framework for describing dense matter in neutron
stars. Magnetised compact-star matter has also been
studied within relativistic mean-field and density-dependent
hadronic models~\cite{rabhi2008stellar}. However, many
treatments of magnetised matter either neglect finite-temperature
effects, omit leptonic consistency, or mix static magnetic-field
effects with fully dynamical magnetic-field evolution. However, many treatments of magnetised matter either neglect finite-temperature effects, omit leptonic consistency, or mix static magnetic-field effects with fully dynamical magnetic-field evolution. In the present work, we restrict attention to a minimal and internally consistent baseline problem: uniform \(npe\mu\) matter in a prescribed external magnetic field, described within a relativistic mean-field framework with \(\sigma\), \(\omega\), and \(\rho\) mesons. In the present work, we consider the linear QHD-I model as a benchmark baseline together with the nonlinear GM1 parameterisation as a more realistic comparison.

The goal of this paper is not to solve the full magneto-thermal evolution problem of neutron stars. Rather, we develop a self-consistent baseline framework for magnetised dense matter that can later be extended to include anomalous magnetic moments, hyperons, quark degrees of freedom, and explicit magnetic-field evolution. This baseline already captures the main effects of Landau quantisation on the composition, energy density, and anisotropic pressure of dense matter. The purpose of this work is to isolate the hierarchy of static magnetic-field effects in a minimal RMF \(npe\mu\) baseline. Rather than introducing additional exotic degrees of freedom or a dynamical magnetic-field profile, we quantify how Landau quantisation, Maxwell pressure anisotropy, and degenerate thermal corrections enter separately, and compare their magnitude with the model dependence between QHD-I and GM1. This allows us to identify the field-strength regime in which the bulk EoS remains essentially unmodified and the regime in which strongly quantising effects begin to appear.

The paper is organised as follows. In Sec.~II, we present the RMF formalism for magnetised $npe\mu$ matter, including the zero-temperature and finite-temperature density expressions, equilibrium conditions, and thermodynamic quantities. In Sec.~III, we describe the self-consistent numerical procedure and present representative results for the density response, equation of state, model dependence, and exploratory stellar-structure implications.

\section{Theory}

\subsection{RMF Lagrangian in a magnetic field}

We use natural units,
\begin{equation}
\hbar=c=k_B=1.
\end{equation}
The metric convention is
\begin{equation}
g_{\mu\nu}=\mathrm{diag}(+,-,-,-).
\end{equation}
The magnetic field is taken to be uniform and oriented along the \(z\)-axis,
\begin{equation}
\mathbf B = B\hat{\mathbf z}.
\end{equation}
The particle charges are
\begin{equation}
q_p=+e,\qquad q_e=q_\mu=-e,\qquad q_n=0,
\end{equation}
which fixes the sign convention used in the covariant derivative
\begin{equation}
D_i^\mu=\partial^\mu+i q_i A^\mu .
\end{equation}
The total Lagrangian density is introduced now as
\begin{equation}
\mathcal{L}=\mathcal{L}_{b}+\mathcal{L}_{\ell}+\mathcal{L}_{m}+\mathcal{L}_{\mathrm{EM}},
\end{equation}
where $\mathcal{L}_{b}$, $\mathcal{L}_{\ell}$, $\mathcal{L}_{m}$, and
$\mathcal{L}_{\mathrm{EM}}$ denote the baryon, lepton, meson, and
electromagnetic contributions, respectively. Since neutron magnetic response is ignored in the present baseline, no anomalous magnetic-moment term of the form 
$\bar{\psi}_i \sigma_{\mu \nu} F^{\mu \nu} \psi_i$
is included. 

Although the electromagnetic Lagrangian is included, Maxwell's equations are not solved self-consistently for a stellar magnetic-field profile. In this baseline EoS study, the magnetic field is treated as a prescribed uniform background field. Its Maxwell contribution to the energy density and pressure anisotropy is included, but the field itself is not dynamically evolved.

For the baryons $b=n,p$, we write
\begin{align}
\mathcal{L}_{b}
= {}&
\sum_{b=n,p}
\bar{\psi}_{b}
[
i\gamma_{\mu}D_{b}^{\mu}
- m_{b} \notag\\
&\quad
+ g_{\sigma b}\sigma
- g_{\omega b}\gamma_{\mu}\omega^{\mu}
- g_{\rho b}\gamma_{\mu}I_{3b}\rho^{\mu}_{3}
]\psi_{b}.
\end{align}

where $\psi_b$ is the baryon spinor, $m_b$ is the bare baryon mass,
$g_{\sigma b}$, $g_{\omega b}$, and $g_{\rho b}$ are the baryon--meson
coupling constants, and $I_{3b}$ denotes the third component of isospin.
Throughout this work we use the isospin convention
\begin{equation}
I_{3p}=+1,\qquad I_{3n}=-1 .
\label{I3pconvension}
\end{equation}

For leptons $\ell=e,\mu$, the Lagrangian is
\begin{equation}
\mathcal{L}_{\ell}
=
\sum_{\ell=e,\mu}
\bar{\psi}_{\ell}
\left(
i\gamma_{\mu}D_{\ell}^{\mu}
- m_{\ell}
\right)
\psi_{\ell}.
\end{equation}
The covariant derivative is defined in Eq.~(5), where \(q_i\) is the electric charge of species \(i\) and \(A^\mu\) is the electromagnetic four-potential.

The mesonic contribution is
\begin{align}
\mathcal{L}_{m}
&=
\frac{1}{2}\partial_{\mu}\sigma\,\partial^{\mu}\sigma
-\frac{1}{2}m_{\sigma}^{2}\sigma^{2}
-U(\sigma)
\nonumber\\
&\quad
-\frac{1}{4}\omega_{\mu\nu}\omega^{\mu\nu}
+\frac{1}{2}m_{\omega}^{2}\omega_{\mu}\omega^{\mu}
\nonumber\\
&\quad
-\frac{1}{4}\rho_{\mu\nu}\rho^{\mu\nu}
+\frac{1}{2}m_{\rho}^{2}\rho_{\mu}\rho^{\mu},
\end{align}
where
\begin{equation}
\omega_{\mu\nu}=\partial_{\mu}\omega_{\nu}-\partial_{\nu}\omega_{\mu},
\qquad
\rho_{\mu\nu}=\partial_{\mu}\rho_{\nu}-\partial_{\nu}\rho_{\mu}.
\end{equation}

The scalar self-interaction potential is taken as
\begin{equation}
U(\sigma)
=
\frac{1}{3}\,b\,m_N\,(g_{\sigma}\sigma)^3
+
\frac{1}{4}\,c\,(g_{\sigma}\sigma)^4,
\label{eq:Usigma}
\end{equation}
where $b$ and $c$ are dimensionless nonlinear coupling constants.
For the QHD-I parameterisation one sets
\begin{equation}
b=c=0,
\end{equation}
thereby recovering the linear Walecka model~\cite{serot1986advance}.
For the GM1 parameterisation~\cite{glendenning1991reconciliation}, $b$ and $c$
take the nonzero values listed in Table~\ref{tab:rmf_params}. The nonlinear
terms reduce the nuclear incompressibility from the very stiff QHD-I value
$K\approx 540\,\mathrm{MeV}$ to the more realistic GM1 value
$K\approx 300\,\mathrm{MeV}$.

In the mean-field approximation, the corresponding scalar-field equation becomes
\begin{equation}
m_\sigma^2\bar{\sigma}
+ b\,m_N\,g_\sigma^3\,\bar{\sigma}^2
+ c\,g_\sigma^4\,\bar{\sigma}^3
=
\sum_{b=n,p} g_{\sigma b}\,n_{s,b},
\label{eq:sigma_NL}
\end{equation}
while all other mean-field equations remain unchanged. 
For \(b=c=0\), this scalar-field equation reduces to the familiar linear relation of the linear Walecka model.

The electromagnetic field contribution is
\begin{equation}
\mathcal{L}_{\mathrm{EM}}
=
-\frac{1}{4}F_{\mu\nu}F^{\mu\nu},
\end{equation}
with
\begin{equation}
F_{\mu\nu}=\partial_{\mu}A_{\nu}-\partial_{\nu}A_{\mu}.
\end{equation}

\subsection{Mean-field approximation and effective single-particle energies}

In uniform matter, the meson fields are replaced by their mean values,
\begin{equation}
\sigma \rightarrow \bar{\sigma},
\qquad
\omega^\mu \rightarrow \delta^\mu_0\,\bar{\omega}_0,
\qquad
\rho^\mu_3 \rightarrow \delta^\mu_0\,\bar{\rho}_{03}.
\end{equation}
The baryon effective mass is then
\begin{equation}
m_b^* = m_b - g_{\sigma b}\bar{\sigma}.
\end{equation}

The effective chemical potential of a baryon is
\begin{equation}
\mu_b^* = \mu_b - g_{\omega b}\bar{\omega}_0 - g_{\rho b}I_{3b}\bar{\rho}_{03}.
\end{equation}

For neutrons, the single-particle energy is
\begin{equation}
E_n(p)=\sqrt{p^2+m_n^{*2}}.
\end{equation}

For charged species $i=p,e,\mu$, Landau quantisation modifies the transverse motion \cite{harding2006physics, potekhin2015neutron}. Their energy spectrum becomes
\begin{equation}
E_{i,\nu}(p_z)=\sqrt{p_z^2+m_i^{*2}+2\nu |q_i|B},
\qquad \nu=0,1,2,\dots
\label{Espectrum of Landau}
\end{equation}
where
\begin{equation}
m_p^*=m_p-g_{\sigma p}\bar{\sigma},
\qquad
m_e^*=m_e,
\qquad
m_\mu^*=m_\mu.
\end{equation}
The Landau level degeneracy is
\begin{equation}
g_\nu = 2-\delta_{\nu 0}.
\label{Landau level degeneracy}
\end{equation}

The quantity $\left|q_i\right| B$ that appears in Eq. [\ref{Espectrum of Landau}] is different from the magnetic field appearing in the Maxwell energy density $\frac{B^2}{2}$. In the analytical formulas, $B$ denotes the magnetic field in natural Heaviside–Lorentz units. Landau quantisation depends on $\left|q_i\right| B$, whereas the Maxwell contribution is $\frac{B^2}{2}$. When numerical values are quoted in gauss, $B$ is converted to natural units before insertion into these expressions.

The Landau index $\nu$ in Eq.[\ref{Landau level degeneracy}] combines the orbital Landau level and the ordinary Dirac spin degeneracy; in the absence of anomalous magnetic moments, the lowest level has degeneracy one and all higher levels have degeneracy two. In this way, the following expression is verified: 

\begin{equation}
\frac{\left|q_i\right| B}{2 \pi^2} \sum_{\nu=0}^{\nu_{\max }} g_\nu \longrightarrow \frac{1}{\pi^2} \int p_{\perp} d p_{\perp} \quad(B \rightarrow 0)
\end{equation}

\subsection{Densities in a magnetic field at zero temperature}

At zero temperature, the Fermi momentum of a charged species in Landau level $\nu$ is
\begin{equation}
k_{F,i,\nu}
=
\sqrt{\mu_i^{*2}-m_i^{*2}-2\nu |q_i|B},
\end{equation}
and the maximum occupied Landau level is
\begin{equation}
\nu_{\max,i}
=
\left\lfloor
\frac{\mu_i^{*2}-m_i^{*2}}{2|q_i|B}
\right\rfloor.
\end{equation}

The number density of a charged species at $T=0$ is
\begin{equation}
n_i
=
\frac{|q_i|B}{2\pi^2}
\sum_{\nu=0}^{\nu_{\max,i}}
g_\nu\,k_{F,i,\nu},
\qquad i=p,e,\mu.
\end{equation}
For neutrons,
\begin{equation}
k_{F,n}=\sqrt{\mu_n^{*2}-m_n^{*2}},
\end{equation}
and
\begin{equation}
n_n=\frac{k_{F,n}^3}{3\pi^2}.
\end{equation}

For the proton, the scalar density is
\begin{equation}
n_{s,p}
=
\frac{|q_p|B}{2\pi^2}
\sum_{\nu=0}^{\nu_{\max,p}}
g_\nu
\int_0^{k_{F,p,\nu}}
\frac{m_p^*\,dp_z}
{\sqrt{p_z^2+m_p^{*2}+2\nu |q_p|B}} .
\end{equation}
Defining
\begin{equation}
a_{\nu,p}^2
=
m_p^{*2}+2\nu |q_p|B,
\end{equation}
one obtains
\begin{equation}
\int_0^{k_{F,p,\nu}}
\frac{m_p^*\,dp_z}
{\sqrt{p_z^2+a_{\nu,p}^2}}
=
m_p^*
\ln\left[
\frac{
k_{F,p,\nu}
+
\sqrt{k_{F,p,\nu}^2+a_{\nu,p}^2}
}
{a_{\nu,p}}
\right].
\end{equation}
Since
\begin{equation}
\sqrt{k_{F,p,\nu}^2+a_{\nu,p}^2}
=
\mu_p^*,
\end{equation}
the proton scalar density becomes
\begin{equation}
n_{s,p}
=
\frac{|q_p|B\,m_p^*}{2\pi^2}
\sum_{\nu=0}^{\nu_{\max,p}}
g_\nu
\ln\left[
\frac{
k_{F,p,\nu}+\mu_p^*
}
{
\sqrt{m_p^{*2}+2\nu |q_p|B}
}
\right].
\end{equation}

The neutron scalar density is
\begin{equation}
n_{s,n}
=
\frac{m_n^*}{2\pi^2}
\left[
k_{F,n}\mu_n^*
-
m_n^{*2}
\ln\left(
\frac{k_{F,n}+\mu_n^*}{m_n^*}
\right)
\right].
\end{equation}

The total baryon density and scalar density are therefore
\begin{equation}
n_B=n_n+n_p,
\qquad
n_s=n_{s,n}+n_{s,p}.
\end{equation}

\subsection{Formal finite-temperature extension and degenerate limit}
\label{sec:finite_T}

For applications at nonzero temperature, the zero-temperature step functions are replaced by Fermi--Dirac occupation factors. For charged species $i=p,e,\mu$, the number density becomes
\begin{equation}
n_i(T)
=
\frac{|q_i|B}{2\pi^2}
\sum_{\nu=0}^{\infty} g_\nu
\int_0^\infty dp_z
\left[
f_i(E_{i,\nu})-\bar f_i(E_{i,\nu})
\right],
\end{equation}
and the proton scalar density generalises to
\begin{align}
& n_{s,p}(T)
=\\
& \frac{|q_p|B\,m_p^*}{2\pi^2}
\sum_{\nu=0}^{\infty} g_\nu
\int_0^\infty
\frac{dp_z}{E_{p,\nu}}
\left[
f_p(E_{p,\nu})+\bar f_p(E_{p,\nu})
\right],
\end{align}

For neutrons, the finite-temperature number density reads

\begin{equation}
n_n(T)=\frac{1}{\pi^2} \int_0^{\infty} p^2\left[f_n\left(E_n\right)-\bar{f}_n\left(E_n\right)\right] d p
\end{equation}

and 

\begin{equation}
n_{s, n}(T)=\frac{m_n^*}{\pi^2} \int_0^{\infty} \frac{p^2}{E_n}\left[f_n\left(E_n\right)+\bar{f}_n\left(E_n\right)\right] d p ,
\end{equation}
where

\begin{align}
f_j(E) {}& =\frac{1}{1+\exp[(E-\mu_j^*)/T]},\\
\bar{f}_j(E) & =\frac{1}{1+\exp[(E+\mu_j^*)/T]}.
\end{align}
for $j=i \text{\quad or \quad} j=n$.

The expressions above define the formal finite-temperature generalisation of the density integrals. In the numerical results of the present first-pass study, however, the complete finite-temperature RMF system is not solved self-consistently. Instead, the \(T=0\) self-consistent solution is first obtained, and moderate thermal effects are estimated perturbatively in the degenerate regime through the leading-order Sommerfeld expansion.

The opposite sign structure in the number and scalar densities reflects the fact that antiparticles contribute with opposite sign to the conserved number density, but positively to the scalar bilinear. In the density and temperature range relevant to neutron-star matter, antiparticle contributions remain numerically small and the zero-temperature limit is smoothly recovered as $T\to 0$ \cite{prakash1997composition}.

\subsection{Mean-field equations and equilibrium conditions}
\label{sec:mf_eqs}

In the mean-field approximation the meson mean fields satisfy
\begin{equation}
m_\sigma^2\bar{\sigma}
+b\,m_N\,g_{\sigma}^3\,\bar{\sigma}^2
+c\,g_{\sigma}^4\,\bar{\sigma}^3
=
\sum_{b=n,p} g_{\sigma b}\,n_{s,b},
\label{eq:sigma_NL}
\end{equation}
\begin{equation}
m_\omega^2\bar{\omega}_0
=
\sum_{b=n,p} g_{\omega b}\,n_b,
\end{equation}
\begin{equation}
m_\rho^2\bar{\rho}_{03}
=
\sum_{b=n,p} g_{\rho b}\,I_{3b}\,n_b.
\label{mesonM}
\end{equation}

Defining the isovector density as $n_I=n_n-n_p,$ the \(\rho\)-field equation may equivalently be written as

\begin{equation}
m_\rho^2 \bar\rho_{03}
=
g_\rho(n_p-n_n)
=
-g_\rho n_I .
\end{equation}

For QHD-I ($b=c=0$) Eq.~(\ref{eq:sigma_NL}) reduces to the
familiar linear relation $m_\sigma^2\bar{\sigma}=\sum_b g_{\sigma b}
n_{s,b}$.
For GM1 ($b,c\neq 0$) it is a cubic equation in $\bar{\sigma}$,
which is solved numerically by a standard root-find at each
iteration step.

For neutron-star matter, beta equilibrium implies
\begin{equation}
\mu_n=\mu_p+\mu_e,
\qquad
\mu_\mu=\mu_e,
\end{equation}
while charge neutrality requires
\begin{equation}
n_p=n_e+n_\mu.
\end{equation}
These conditions, together with the field equations above,
determine the composition of matter for a given baryon density
and magnetic-field strength.

\subsection{Energy density and pressure}

The total energy density is written as
\begin{equation}
\varepsilon
=
\varepsilon_b+\varepsilon_\ell+\varepsilon_{\mathrm{mes}}+\varepsilon_{\mathrm{EM}},
\end{equation}
where the mesonic and electromagnetic contributions are
\begin{equation}
\varepsilon_{\mathrm{mes}}
=
\frac{1}{2}m_\sigma^2\bar{\sigma}^2
+U(\bar{\sigma})
+\frac{1}{2}m_\omega^2\bar{\omega}_0^2
+\frac{1}{2}m_\rho^2\bar{\rho}_{03}^2,
\label{eq:eps_mes}
\end{equation}
\begin{equation}
\varepsilon_{\mathrm{EM}}=\frac{B^2}{2}.
\end{equation}
For charged species,

\begin{align}
\varepsilon_i
& =
\frac{|q_i|B}{2\pi^2}
\sum_{\nu=0}^{\nu_{\max,i}}
g_\nu
\int_0^{k_{F,i,\nu}}
\sqrt{p_z^2+m_i^{*2}+2\nu |q_i|B}\,dp_z,\\
 i & =p,e,\mu,
  \end{align}
whereas for neutrons,
\begin{equation}
\varepsilon_n
=
\frac{1}{\pi^2}
\int_0^{k_{F,n}}
\sqrt{p^2+m_n^{*2}}\,p^2\,dp.
\end{equation}
Thus,
\begin{equation}
\varepsilon_b=\varepsilon_n+\varepsilon_p,
\qquad
\varepsilon_\ell=\varepsilon_e+\varepsilon_\mu.
\end{equation}

At zero temperature, the matter pressure is obtained from the thermodynamic relation
\begin{equation}
P_{\mathrm{mat}}
=
\sum_{i=n,p,e,\mu}\mu_i n_i
-
\left(
\varepsilon_b+\varepsilon_\ell+\varepsilon_{\mathrm{mes}}
\right).
\end{equation}
This relation is not an independent field equation; rather, it follows from the Euler relation once the mean fields and particle densities have been determined self-consistently.

Because the magnetic field breaks rotational symmetry, the total pressure is anisotropic. Defining the matter pressure by
\begin{equation}
P_{\mathrm{mat}}=-\Omega_{\mathrm{mat}},
\end{equation}
the matter-sector directional pressures satisfy
\begin{equation}
P_{\parallel}^{\mathrm{mat}}=P_{\mathrm{mat}},
\qquad
P_{\perp}^{\mathrm{mat}}=P_{\mathrm{mat}}-MB,
\end{equation}
where
\begin{equation}
M=\left(\frac{\partial P_{\mathrm{mat}}}{\partial B}\right)_{\{\mu_i\},T}
\end{equation}
is the magnetisation. In the present first-pass treatment, magnetisation effects are neglected, $M\simeq 0$. Including only the Maxwell contribution, the total directional pressures are approximated by
\begin{equation}
P_{\parallel}=P_{\mathrm{mat}}-\frac{B^2}{2},
\end{equation}
\begin{equation}
P_{\perp}=P_{\mathrm{mat}}+\frac{B^2}{2}.
\end{equation}

so

\begin{equation}
  \boxed{P_{\mathrm{iso}}=P_{\mathrm{mat}}+\frac{B^2}{6}}
\end{equation}

and with $\varepsilon_{\mathrm{mat}}=\varepsilon_b+\varepsilon_{\ell}+\varepsilon_{\mathrm{mes}}$ the total energy density is

\begin{equation}
  \boxed{\varepsilon_{\mathrm{tot}}=\varepsilon_{\mathrm{mat}}+\frac{B^2}{2}}
\end{equation}
The distinction between longitudinal and transverse pressures follows from the breaking of rotational symmetry by the magnetic field and has been discussed for dense magnetised fermionic systems~\cite{ferrer2010equation,strickland2012bulk}. In the present baseline calculation we retain only the Maxwell contribution to the pressure tensor and neglect the matter magnetisation term \(MB\). Within this no-magnetisation approximation, the magnetic field lowers the longitudinal pressure and raises the transverse pressure. The magnetisation contribution can be evaluated explicitly and may become relevant at sufficiently strong fields~\cite{dong2013magnetization}.

\section{Results and Numerical Implementation}

For fixed baryon density \(n_B\) and magnetic-field strength \(B\), the coupled mean-field equations are solved self-consistently at \(T=0\). Finite-temperature effects shown below are then estimated perturbatively in the degenerate regime using the leading-order Sommerfeld correction. The calculations reported here are performed for charge-neutral, beta-equilibrated $npe\mu$ matter in a uniform external magnetic field, using the RMF framework introduced in Sec.~II. Two parameterisations are considered throughout: the linear QHD-I model as a benchmark baseline, and the nonlinear GM1 model as a more realistic comparison.

Unless otherwise stated, natural units with $\hbar=c=1$ are used throughout. When needed numerically, magnetic fields are converted from gauss to energy-squared units in the standard way. The elementary charge is taken as
\begin{equation}
e=\sqrt{4\pi\alpha}\simeq 0.303.
\end{equation}

For exploratory isotropised-TOV estimates, the anisotropic pressures are defined consistently with Sec.~II as
\begin{equation}
P_\parallel=P_{\rm mat}-\frac{B^2}{2},
\qquad
P_\perp=P_{\rm mat}+\frac{B^2}{2},
\end{equation}
with magnetisation neglected in the present first-pass treatment. We follow the standard convention for directional pressures in magnetised fermionic matter discussed by Strickland, Dexheimer, and Menezes~\cite{strickland2012bulk}. When a scalar pressure proxy is required for approximate, diagnostic TOV integrations, we use
\begin{equation}
P_{\rm iso}=\frac{2P_\perp+P_\parallel}{3},
\end{equation}
while noting that this prescription is exploratory and cannot replace a fully anisotropic Einstein--Maxwell stellar-structure calculation.

In the present implementation, hyperons, deconfined quarks, and anomalous magnetic moments are not included ~\cite{broderick2000equation}. The isovector density is treated self-consistently as
\begin{equation} n_I = n_n - n_p .
\end{equation}
With the convention \(I_{3p}=+1\) and \(I_{3n}=-1\), the \(\rho\)-field equation may be written as
\begin{equation}
m_\rho^2 \bar\rho_{03}
=
g_\rho(n_p-n_n)
=
-g_\rho n_I .
\end{equation}
and no symmetric-matter approximation is imposed.
The numerical calculations are performed using the two
parameterisations listed in Table~\ref{tab:rmf_params}:
the linear QHD-I model~\cite{serot1986advance} as a
minimal baseline, and the nonlinear GM1
model~\cite{glendenning1991reconciliation} as a more realistic
comparison. In QHD-I the nonlinear couplings are set
to $b = c = 0$; in GM1 they take the values given in
Table~\ref{tab:rmf_params} and the scalar field
equation is solved as the cubic
Eq.~(\ref{eq:sigma_NL}) at each iteration step.The bare nucleon masses are $m_n = 939.6\,\mathrm{MeV}$ and $m_p = 938.3\,\mathrm{MeV}$; lepton masses are $m_e = 0.511\,\mathrm{MeV}$ and $m_\mu = 105.7\,\mathrm{MeV}$. The saturation properties reproduced by each set are listed in Table~\ref{tab:rmf_params}.

\begin{table}[h]
\centering
\caption{RMF coupling constants and meson masses for the
two parameterisations used in this work.
\textit{QHD-I}: linear Walecka model~\cite{serot1986advance},
used as a minimal baseline.
\textit{GM1}: nonlinear Glendenning--Moszkowski
model~\cite{glendenning1991reconciliation}, used as a more realistic
comparison.
Both sets use $m_N = 939\,\mathrm{MeV}$,
$m_e = 0.511\,\mathrm{MeV}$,
$m_\mu = 105.7\,\mathrm{MeV}$.
The lower block lists the symmetric nuclear matter
saturation properties reproduced by each set.}
\label{tab:rmf_params}
\begin{tabular}{lcc}
\hline\hline
Quantity & QHD-I & GM1 \\
\hline
$g_\sigma$       & 10.075 & 8.910  \\
$g_\omega$       & 13.200 & 10.610 \\
$g_\rho$         &  4.000 &  4.093 \\
$m_\sigma$ [MeV] &  520.0 & 512.0  \\
$m_\omega$ [MeV] &  783.0 & 783.0  \\
$m_\rho$   [MeV] &  770.0 & 770.0  \\
$b$              &  0     & 0.002947 \\
$c$              &  0     & $-$0.001070 \\
\hline
\multicolumn{3}{l}{\textit{Saturation properties (SNM):}} \\
\hline
$n_0\,[\mathrm{fm}^{-3}]$  & 0.160 & 0.153 \\
$E/A - m_N\,[\mathrm{MeV}]$& $-$16.0 & $-$16.3 \\
$m^*/m_N$                  & 0.543 & 0.700 \\
$K\,[\mathrm{MeV}]$        & $\approx$540 & $\approx$300 \\
$J_\mathrm{sym}\,[\mathrm{MeV}]$ & $\approx$38 & 32.5 \\
\hline\hline
\end{tabular}
\end{table}
\subsection{Self-consistent numerical procedure}

For each chosen pair \((n_B,B)\), the \(T=0\) coupled system is solved iteratively as follows:
\begin{enumerate}
    \item Initialize the mean fields $\bar{\sigma}$, $\bar{\omega}_0$, and $\bar{\rho}_{03}$ together with trial chemical potentials.
    
    \item Compute the effective masses and effective chemical potentials using the relations introduced in Sec.~II.
    
    \item Evaluate the particle number densities and scalar densities for neutrons, protons, electrons, and muons.
    
    \item Update the mean fields using the RMF field equations,
    \begin{align}
    m_\sigma^2 \bar{\sigma}
    + b\,m_N\,g_\sigma^3\,\bar{\sigma}^2
    + c\,g_\sigma^4\,\bar{\sigma}^3
    &= \sum_{b=n,p} g_{\sigma b} n_{s,b},
    \\
    m_\omega^2 \bar{\omega}_0
    &= \sum_{b=n,p} g_{\omega b} n_b,
    \\
    m_\rho^2 \bar{\rho}_{03}
    &= \sum_{b=n,p} g_{\rho b} I_{3b} n_b.
    \end{align}
    For the QHD-I parameterisation, the nonlinear couplings vanish,
    $b=c=0$, and the scalar-field equation reduces to the usual linear form.
    For the GM1 parameterisation, the scalar-field equation is nonlinear and
    is solved numerically at each iteration step.
    
    \item Enforce beta equilibrium and charge neutrality,
    \begin{equation}
    \mu_n=\mu_p+\mu_e,
    \qquad
    \mu_\mu=\mu_e,
    \qquad
    n_p=n_e+n_\mu.
    \end{equation}
    
    \item Repeat steps 2--5 until the relative changes in all mean fields and chemical potentials fall below the chosen convergence tolerance.
    
    \item After convergence, compute the total energy density, matter pressure, and directional pressures $(P_{\parallel},P_{\perp})$.
\end{enumerate}
When finite-temperature corrections are displayed, they are added after convergence of the \(T=0\) solution using the leading-order Sommerfeld expression for the thermal pressure increment.

\subsection{Density response to magnetic field}
\label{sec:density_response}

As a baseline validation, the $B=0$ limit of our implementation
reproduces the QHD-I saturation properties listed in
Table~\ref{tab:rmf_params}: saturation density
$n_0^{\rm QHD-I}=0.193\,\mathrm{fm}^{-3}$, binding energy
$E/A-m_N=-15.75\,\mathrm{MeV}$, and effective-mass ratio
$m^*/m_N=0.54$, in agreement with the original linear
Walecka model~\cite{serot1986advance,walecka1974theory}.
Throughout the figures, however, densities are normalised to the
conventional empirical saturation density
$n_0^{\rm emp}=0.16\,\mathrm{fm}^{-3}$, following common
practice in the RMF literature.
In $\beta$-equilibrated matter at $B=0$, the proton fraction
reaches $x_p \approx 0.04$ near saturation, consistent with
previous linear RMF calculations~\cite{glendenning2012compact}.

Table~\ref{tab:eos} lists representative thermodynamic quantities
for the $B=0$, $T=0$ baseline. At zero magnetic field the pressure
is isotropic, so that
\begin{equation}
P_\parallel=P_\perp=P_{\rm iso}=P_{\rm mat}.
\end{equation}
The full machine-readable table, including directional pressures
and finite-temperature results for all $(B,T)$ combinations
considered in this work, is provided as supplementary material.

{\small
\begin{table*}[t]
\centering
\caption{Representative equation-of-state table for
$\beta$-equilibrated $n$-$p$-$e$-$\mu$ matter at $B=0$
and $T=0$, obtained from the self-consistent linear RMF
calculation with the parameters of
Table~\ref{tab:rmf_params}. Columns list the baryon density
$n_B$ in units of the empirical saturation density
$n_0^{\rm emp}=0.16\,\mathrm{fm}^{-3}$, the total energy density
$\varepsilon$, the isotropised pressure $P_{\rm iso}$, the proton
fraction $x_p=n_p/n_B$, the neutron effective-mass ratio
$m_n^*/m_N$, and the neutron and electron chemical potentials
$\mu_n$ and $\mu_e$.}
\label{tab:eos}
\begin{tabular}{crrcccc}
\hline\hline
$n_B/n_0$ &
$\varepsilon\;[\mathrm{MeV\,fm}^{-3}]$ &
$P_{\rm iso}\;[\mathrm{MeV\,fm}^{-3}]$ &
$x_p$ &
$m_n^*/m_N$ &
$\mu_n\;[\mathrm{MeV}]$ &
$\mu_e\;[\mathrm{MeV}]$ \\
\hline
$0.5$ &  $77.2$ &    $0.07$ & $0.009$ & $0.758$ &  $942.8$ &  $55.0$ \\
$1.0$ & $152.7$ &    $4.29$ & $0.029$ & $0.554$ &  $974.8$ & $102.3$ \\
$1.5$ & $232.9$ &   $22.32$ & $0.058$ & $0.400$ & $1062.8$ & $143.1$ \\
$2.0$ & $322.2$ &   $61.63$ & $0.093$ & $0.303$ & $1202.4$ & $173.0$ \\
$2.5$ & $423.9$ &  $122.36$ & $0.131$ & $0.244$ & $1371.3$ & $194.8$ \\
$3.0$ & $539.6$ &  $201.94$ & $0.170$ & $0.206$ & $1552.9$ & $211.9$ \\
$4.0$ & $816.4$ &  $431.50$ & $0.256$ & $0.339$ & $1990.6$ & $178.5$ \\
$5.0$ &$1171.9$ &  $716.23$ & $0.319$ & $0.269$ & $2369.4$ & $218.7$ \\
$6.0$ &$1568.6$ & $1045.88$ & $0.371$ & $0.224$ & $2733.7$ & $249.8$ \\
\hline\hline
\end{tabular}
\end{table*}
}

\begin{figure*}[t]
    \includegraphics[width=0.95\textwidth]{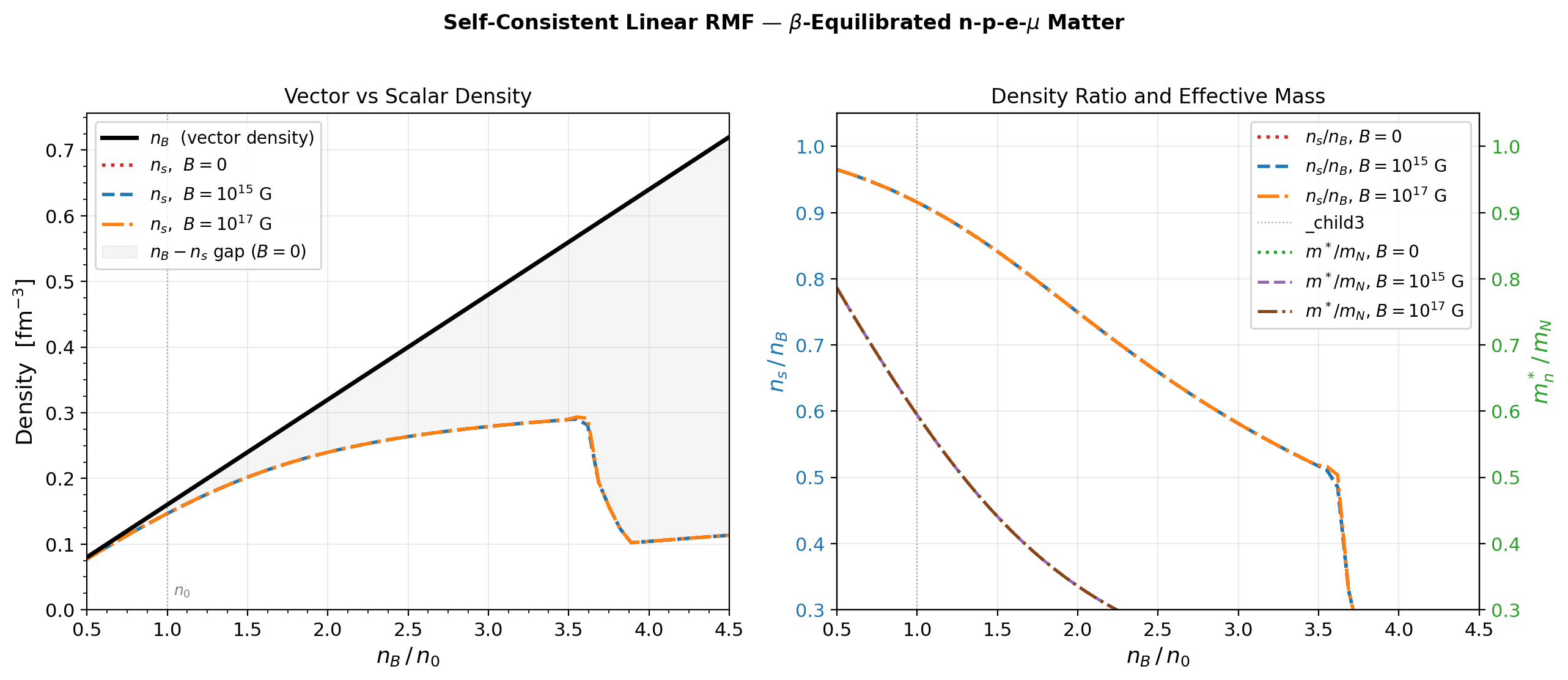}
    \caption{Vector density $n_B$ and scalar density $n_s$ as functions
    of baryon density $n_B/n_0$ for $\beta$-equilibrated neutron-star
    matter ($n$-$p$-$e$-$\mu$ composition) obtained from a self-consistent
    linear RMF calculation with the parameters of
    Table~\ref{tab:rmf_params}. The horizontal axis is normalised to the
    conventional empirical saturation density
    $n_0^{\rm emp}=0.16\,\mathrm{fm}^{-3}$, while the QHD-I model itself
    saturates at $0.193\,\mathrm{fm}^{-3}$.
    \textit{Left panel:} $n_B$ (solid black) and $n_s$ for
    $B=0$ (dotted red), $B=10^{15}\,\mathrm{G}$ (dashed blue), and
    $B=10^{17}\,\mathrm{G}$ (dot-dashed orange). The shaded region
    illustrates the relativistic suppression $n_s<n_B$, which arises
    from the kinematic factor $M^*/E^*<1$ and grows with density as the
    effective mass decreases.
    \textit{Right panel:} Ratio $n_s/n_B$ (left axis) and neutron
    effective-mass ratio $m_n^*/m_N$ (right axis) for the same field
    strengths. At $B=10^{15}\,\mathrm{G}$, the discrete Landau levels are
    sufficiently dense ($\nu_{\max}\sim10^4$) to reproduce the continuum
    $B=0$ result; visible departures from the unmagnetised limit emerge
    only above $B\sim10^{17}\,\mathrm{G}$, where fewer levels are occupied
    and lowest-Landau-level effects begin to appear.}
    \label{fig:densities_vs_nB}
\end{figure*}

Figure~\ref{fig:densities_vs_nB} shows the behaviour of the vector
and scalar densities in the magnetised RMF framework. The charged
species are evaluated using the Landau-level cutoff
\begin{equation}
\nu_{\max,i}
=
\left\lfloor
\frac{\mu_i^{*2}-m_i^{*2}}{2|q_i|B}
\right\rfloor,
\end{equation}
together with the corresponding longitudinal Fermi momenta
\begin{equation}
k_{F,i,\nu}
=
\sqrt{\mu_i^{*2}-m_i^{*2}-2\nu|q_i|B}.
\end{equation}
As the magnetic field increases, the spacing between Landau
levels widens and the number of occupied levels decreases. This
modifies the density of states and produces non-smooth behaviour
in charged-particle observables whenever the highest occupied
Landau level changes. Within RMF theory, the scalar density
remains smaller than the corresponding vector density because it
carries the relativistic weighting factor $m_i^*/E_{i,\nu}$.

The left panel of Fig.~\ref{fig:densities_vs_nB} shows that the
scalar density $n_s$ is suppressed below the vector density $n_B$
by approximately $10\%$ at $n_0$ and by $30$--$40\%$ at
$4\,n_0$, reflecting the relativistic factor $M^*/E^*<1$ that
grows as the effective mass $m^*$ decreases with density. This
suppression is physically significant because $n_s$ drives the
scalar mean field $\bar{\sigma}$ and hence the effective mass
itself; neglecting it would overestimate $m^*$ and underestimate
the stiffness of the equation of state at supranuclear densities.

At $B=10^{15}\,\mathrm{G}$ the curves are effectively
indistinguishable from the $B=0$ result across the entire density
range shown, confirming that Landau quantisation effects on bulk
thermodynamic quantities are negligible at this field strength.
A clear departure appears only at $B=10^{17}\,\mathrm{G}$, where
the effective-mass ratio $m^*/m_N$ develops a visible kink near
$n_B/n_0 \approx 3.5$, associated with a change in the maximum
occupied proton Landau level $\nu_{\max}$. The right panel further
shows that the ratio $n_s/n_B$ decreases from approximately $0.90$
at $n_0$ to $0.60$ at $4\,n_0$, while remaining essentially
independent of $B$ at $10^{15}\,\mathrm{G}$. Together, these
results provide a clean demonstration of the relativistic
many-body suppression expected in RMF matter and of the fact that
magnetic-field effects on bulk densities remain weak until the
system enters the strongly quantising regime.

\subsection{Exploratory isotropised-TOV sensitivity estimates}
\label{sec:tov}

To obtain a first diagnostic estimate of how the magnetised equation of state could affect ordinary spherical mass--radius sequences, we perform exploratory Tolman--Oppenheimer--Volkoff (TOV) integrations using the isotropised pressure
\begin{equation}
P_{\rm iso}=P_{\rm mat}+\frac{B^2}{6}.
\end{equation}
The quantity $P_{\mathrm{iso}}$ is used only as a scalar diagnostic and for exploratory comparison with ordinary TOV solutions. It is not derived from a self-consistent anisotropic Einstein–Maxwell stellar model.

This calculation should not be interpreted as a fully self-consistent model of a strongly magnetised star. The ordinary TOV equations assume an isotropic perfect fluid, whereas the magnetic field produces a pressure tensor with distinct longitudinal and transverse components. The use of \(P_{\rm iso}\) is therefore only a scalar proxy designed to estimate the size of the magnetic-field correction in a familiar spherical-TOV framework.
In geometrised units $G=c=1$, the stellar structure equations are
\begin{equation}
\frac{dP}{dr}
=
-
\frac{(\varepsilon+P)\left(M+4\pi r^3 P\right)}
{r(r-2M)},
\qquad
\frac{dM}{dr}=4\pi r^2 \varepsilon,
\end{equation}
where \(M(r)\) is the enclosed gravitational mass. The isotropised TOV calculation is used only as a diagnostic proxy. It is expected to be least problematic when the anisotropy parameter remains small at neutron-star core densities, but it is not a substitute for a fully anisotropic Einstein--Maxwell stellar-structure calculation.

\begin{figure*}[t]
    \centering
    \includegraphics[width=\textwidth]{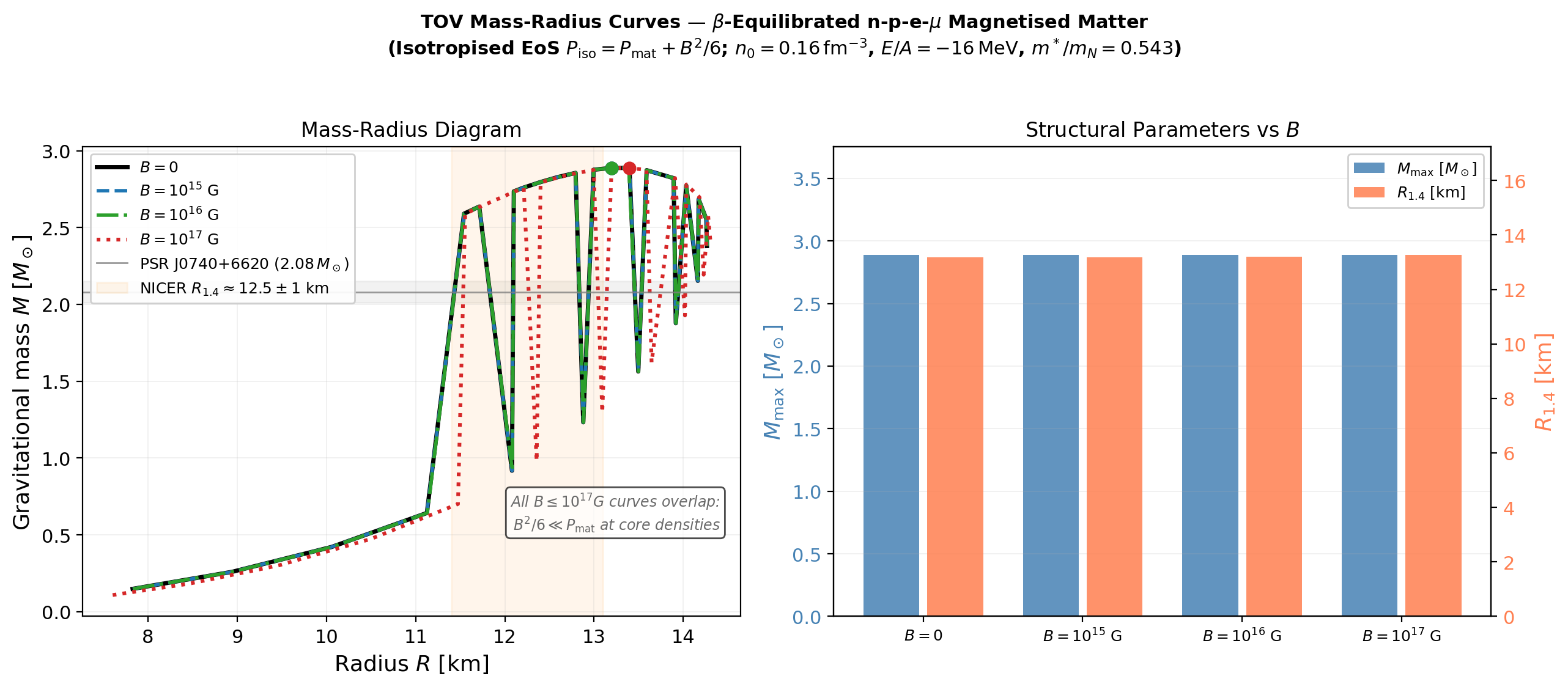}
    \caption{Tolman--Oppenheimer--Volkoff mass-radius
    curves for $\beta$-equilibrated $n$-$p$-$e$-$\mu$
    matter at $T = 0$, computed from the isotropised
    equation of state $P_\mathrm{iso} =
    P_\mathrm{mat} + B^2/6$ for $B = 0$ (solid black),
    $10^{15}$\,G (dashed blue), $10^{16}$\,G
    (dot-dashed green), and $10^{17}$\,G (dotted red).
    Filled circles mark the maximum-mass configuration
    for each field strength.
    \textit{Left panel:} Mass-radius diagram.
    The grey band and horizontal line indicate the
    $2.08\,M_\odot$ mass constraint from
    PSR~J0740+6620~\cite{fonseca2021refined}; the orange band
    shows the NICER radius estimate for a
    $1.4\,M_\odot$ star~\cite{miller2021radius}.
    All curves overlap to within line width for
    $B \leq 10^{16}$\,G; only $B = 10^{17}$\,G
    produces a visible shift ($\Delta R_{1.4} < 0.1$\,km).
    \textit{Right panel:} Maximum mass
    $M_\mathrm{max}$ and radius $R_{1.4}$ as functions
    of field strength, showing that both are insensitive
    to $B$ at astrophysically relevant magnetar field
    strengths.
   The isotropised TOV treatment is used here only as a diagnostic proxy; its use is expected to be least problematic when the anisotropy parameter satisfies \(\Delta\ll1\) at neutron-star core densities
    (see Fig.~\ref{fig:anisotropy}).}
    \label{fig:tov}
\end{figure*}

Figure~\ref{fig:tov} shows the resulting mass-radius
curves for $B = 0$, $10^{15}$, $10^{16}$, and
$10^{17}$\,G at $T = 0$.
All four curves are nearly indistinguishable.
The Maxwell contribution to the isotropised pressure,
$B^2/6$, amounts to $8 \times 10^{-6}$,
$8 \times 10^{-4}$, and $0.08\,\mathrm{MeV\,fm}^{-3}$
at $B = 10^{15}$, $10^{16}$, and $10^{17}$\,G
respectively, compared with $P_\mathrm{mat}(n_0)
\approx 4.3\,\mathrm{MeV\,fm}^{-3}$.
Even at $B = 10^{17}$\,G the fractional shift in
$P_\mathrm{iso}$ is below $2\%$ at $n_0$, and smaller
still at higher densities where the matter pressure
dominates.
Consequently, the maximum mass and the radius at
$1.4\,M_\odot$ are unchanged to three significant figures
for $B \leq 10^{16}$\,G, and shift by less than
$0.1\%$ at $B = 10^{17}$\,G (Table~\ref{tab:tov}).

\begin{table}[h]
\centering
\caption{Maximum gravitational mass $M_\mathrm{max}$,
radius at the maximum-mass configuration
$R(M_\mathrm{max})$, and radius at
$1.4\,M_\odot$ for each magnetic-field strength.
The isotropised TOV approximation is used throughout;
see text for caveats.}
\label{tab:tov}
\begin{tabular}{lccc}
\hline\hline
$B$ &
$M_\mathrm{max}\;[M_\odot]$ &
$R(M_\mathrm{max})\;[\mathrm{km}]$ &
$R_{1.4}\;[\mathrm{km}]$ \\
\hline
$0$            & 2.888 & 13.20 & 13.19 \\
$10^{15}$\,G   & 2.888 & 13.20 & 13.19 \\
$10^{16}$\,G   & 2.888 & 13.20 & 13.19 \\
$10^{17}$\,G   & 2.888 & 13.39 & 13.27 \\
\hline\hline
\end{tabular}
\end{table}

The maximum mass $M_{\max}\approx 2.89\,M_\odot$ exceeds the current observational lower bound from PSR~J0740+6620~\cite{fonseca2021refined}, consistent with the known stiffness of the linear Walecka model. The radius $R_{1.4}\approx 13.2\,\mathrm{km}$~\cite{miller2021radius}, lies somewhat above the most favoured observational range, again reflecting the large incompressibility of the linear model adopted here. Within the isotropised-TOV approximation, the main conclusion is therefore not a precision statement about realistic neutron-star structure, but a robustness statement: for $B\lesssim 10^{17}\,\mathrm{G}$, magnetic-field contributions remain too small to induce appreciable changes in the bulk mass--radius relation. The present TOV results should thus be interpreted as exploratory structural implications of the baseline magnetised EoS, rather than as a fully anisotropic magnetised-star calculation.

\subsection{Equation of state as a function of magnetic field and temperature}
Using the converged \(T=0\) self-consistent RMF solutions, we evaluate the total zero-temperature energy density as
\begin{equation}
\epsilon_{\rm tot}^{(0)}(B)
=
\epsilon_b^{(0)}(B)
+
\epsilon_\ell^{(0)}(B)
+
\epsilon_{\rm mes}^{(0)}(B)
+
\frac{B^2}{2}.
\end{equation}

Here the baryonic and leptonic contributions include the Landau-quantised charged species and the neutron sector. Finite-temperature effects discussed below are included only through the perturbative Sommerfeld correction to the pressure.
\begin{figure*}[t]
    \centering
    \includegraphics[width=0.95\textwidth]{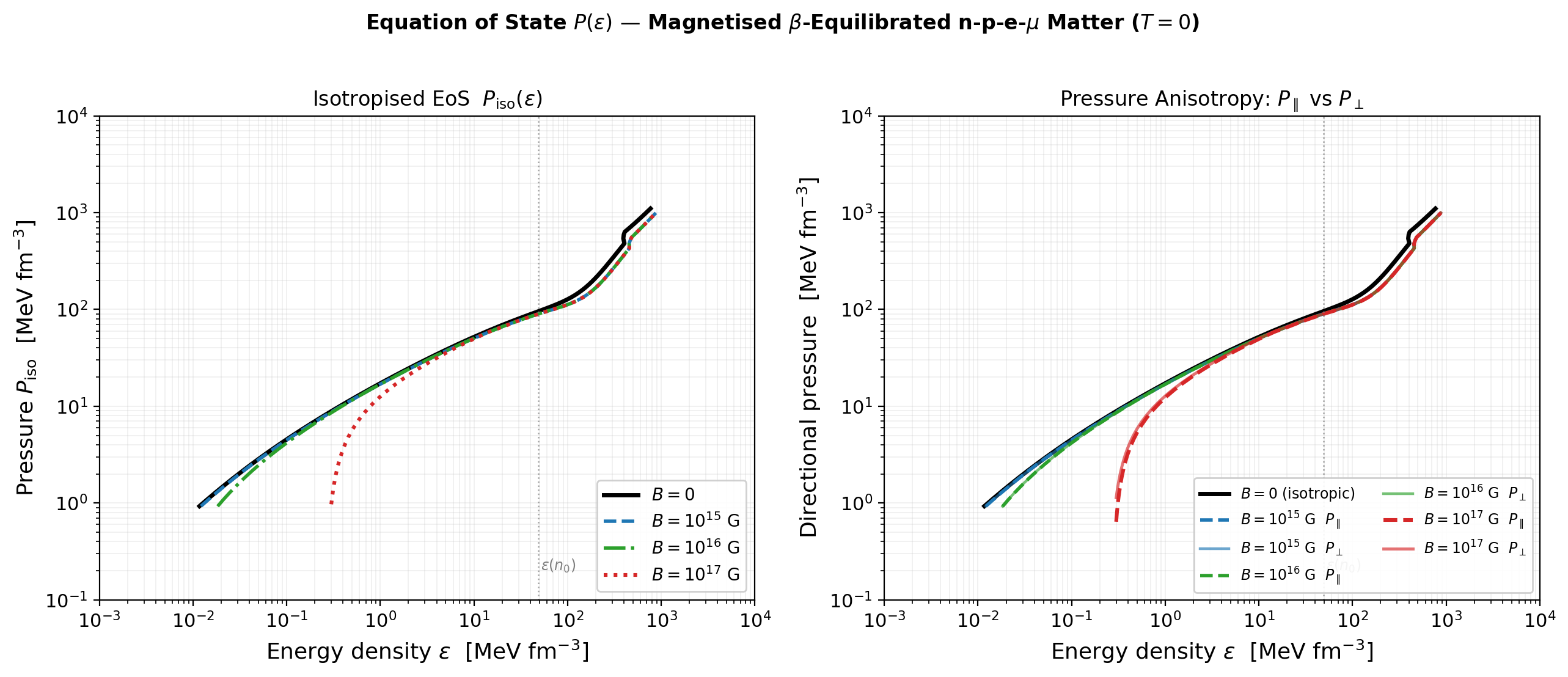}
    \caption{Equation of state for magnetised $\beta$-equilibrated $n$-$p$-$e$-$\mu$ matter at $T=0$ for magnetic-field strengths from $B=0$ to $10^{17}\,\mathrm{G}$. \textit{Left panel:} isotropised pressure $P_{\mathrm{iso}}(\varepsilon)$. \textit{Right panel:} directional pressures $P_{\parallel}$ and $P_{\perp}$, showing the small pressure splitting induced by the magnetic field. The bulk equation of state remains only weakly modified up to $B=10^{17}\,\mathrm{G}$.}
    \label{fig:eos_B}
\end{figure*}

Figure~\ref{fig:eos_B} shows the isotropised pressure
$P_\mathrm{iso}(\varepsilon)$ and the directional pressures
$P_\parallel$, $P_\perp$ for $B = 0$ to $10^{17}$\,G at
$T = 0$. In the left panel, all curves coincide below
$\varepsilon \sim 10^2\,\mathrm{MeV\,fm}^{-3}$, confirming
that the Maxwell contribution $B^2/2$ is negligible relative
to the matter pressure at densities approaching and exceeding
$n_0$; the field contributes at most a few percent to
$P_\mathrm{iso}$ even at $B = 10^{17}$\,G in this density
regime. The right panel makes the anisotropy directly visible:
at $B = 10^{17}$\,G the splitting between $P_\parallel$ and
$P_\perp$ reaches $\Delta P \sim B^2/2 \approx
10^{-2}\,\mathrm{MeV\,fm}^{-3}$ at low energy density,
where the matter pressure is comparably small, but becomes
a sub-percent effect at $\varepsilon \gtrsim
10^2\,\mathrm{MeV\,fm}^{-3}$. For \(B\leq 10^{15}\,\mathrm{G}\), the pressure splitting is negligible at densities relevant to neutron-star interiors. This suggests that an isotropic EoS provides a useful first approximation to the bulk response at canonical magnetar field strengths, although a fully anisotropic treatment would still be required for a complete magnetised-star calculation. The visible stiffening at high energy density in
this baseline calculation is primarily a feature of the stiff
QHD-I parameterisation and is not driven by the magnetic
field itself.

\paragraph{Model dependence: QHD-I versus GM1.}

\begin{figure*}[t]
    \centering
    \includegraphics[width=0.95\textwidth]{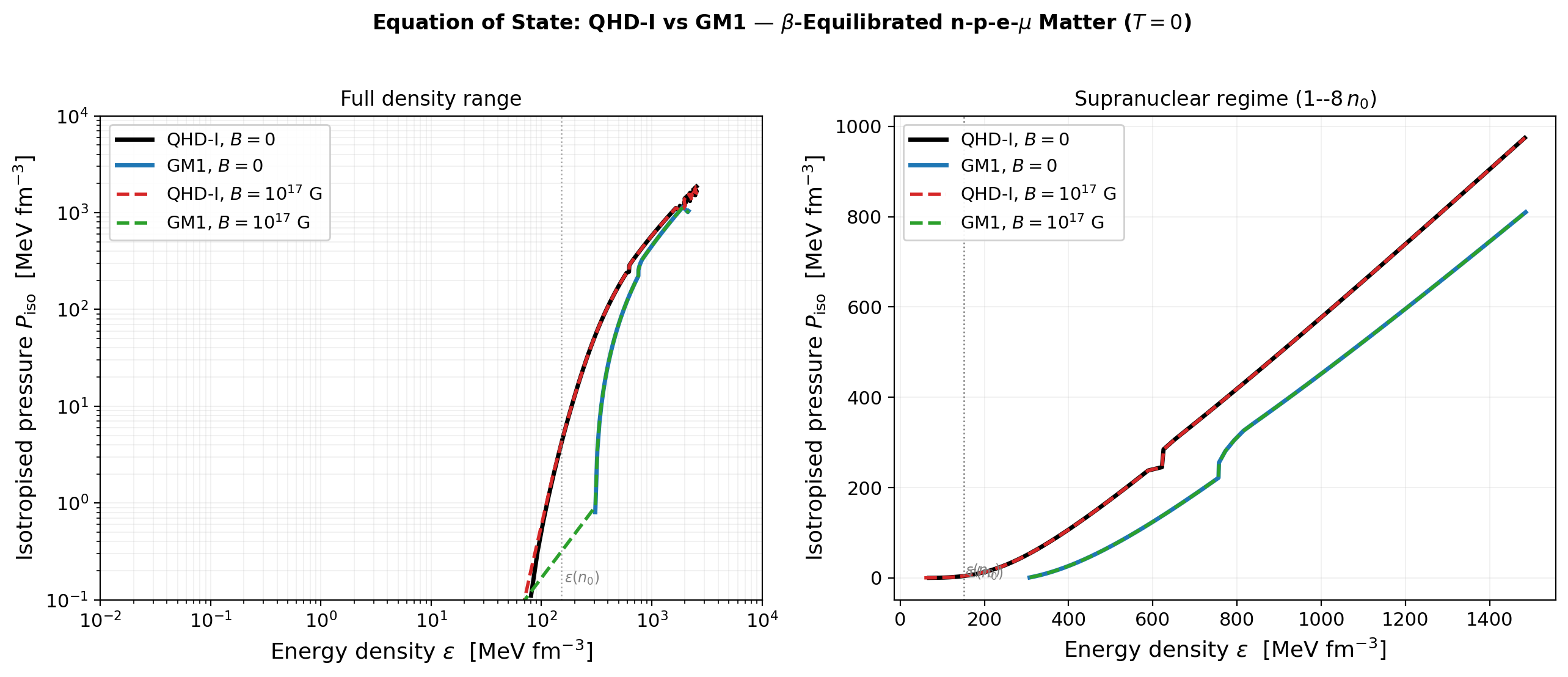}
    \caption{Comparison of the isotropised equation of state $P_{\mathrm{iso}}(\varepsilon)$ for the QHD-I and GM1 parameterisations in $\beta$-equilibrated $n$-$p$-$e$-$\mu$ matter at $T=0$. Solid curves show the unmagnetised limit, while dashed curves correspond to $B=10^{17}\,\mathrm{G}$. \textit{Left panel:} full density range on logarithmic axes. \textit{Right panel:} enlarged view of the supranuclear regime. The GM1 equation of state is systematically softer than QHD-I, reflecting its lower incompressibility and more realistic saturation properties. In contrast, the magnetic-field correction remains very small in both parameterisations up to $B=10^{17}\,\mathrm{G}$, indicating that the weak sensitivity of the bulk equation of state to magnetic field is not an artefact of the stiff linear Walecka baseline.}
    \label{fig:eos_qhdi_gm1}
\end{figure*}

To assess the extent to which the magnetic-field response depends on the choice of RMF parameterisation, Fig.~\ref{fig:eos_qhdi_gm1} compares the isotropised equation of state obtained from the QHD-I and GM1 models at $T=0$, both in the unmagnetised limit and at $B=10^{17}\,\mathrm{G}$. As expected, the GM1 parameterisation yields a systematically softer equation of state than QHD-I over the full supranuclear density range, reflecting the lower incompressibility produced by the nonlinear scalar self-interactions. The difference between the two parameterisations is significantly larger than the magnetic-field correction itself: in both models, the $B=10^{17}\,\mathrm{G}$ curve remains very close to the corresponding $B=0$ result.

This comparison is important for the interpretation of the present results. The linear Walecka model is known to produce an equation of state that is too stiff at high density, and conclusions drawn from QHD-I alone could therefore be questioned as model dependent. Figure~\ref{fig:eos_qhdi_gm1} shows, however, that the weak sensitivity of the bulk equation of state to magnetic field is not an artefact of the stiff linear baseline. Rather, the same qualitative behaviour persists in the more realistic GM1 model: the absolute stiffness of the equation of state changes substantially, but the effect of magnetic fields up to $B=10^{17}\,\mathrm{G}$ remains modest in both cases.

The comparison in Fig.~\ref{fig:eos_qhdi_gm1} therefore provides a useful robustness check on the main conclusion of this section: although the absolute stiffness of the equation of state is strongly model dependent, the magnetic correction to the bulk thermodynamics remains subdominant up to at least $B\sim10^{17}\,\mathrm{G}$. This strengthens the interpretation that the onset of strong Landau-quantisation effects in the bulk equation of state occurs only at still larger fields, even when a more realistic RMF parameterisation is used.

At \(T=0\), the matter pressure used in the equation of state is computed from the thermodynamic relation
\begin{equation}
P_{\rm mat}^{(0)}(B)
=
\sum_{i=n,p,e,\mu}\mu_i n_i
-
\epsilon_{\rm mat}^{(0)}(B),
\end{equation}
where
\begin{equation}
\epsilon_{\rm mat}^{(0)}
=
\epsilon_b+\epsilon_\ell+\epsilon_{\rm mes}.
\end{equation}
This expression is evaluated on the self-consistent \(T=0\) mean-field solution.

At finite temperature, a full thermodynamic treatment would require
\begin{equation}
P = Ts+\sum_i \mu_i n_i-\epsilon,
\end{equation}
or equivalently \(P=-\Omega\), evaluated on a self-consistent finite-temperature solution. In the present work, finite-temperature effects are instead added perturbatively in the degenerate regime through
\begin{equation}
P_{\rm iso}(B,T)
\simeq
P_{\rm iso}^{(0)}(B)+\delta P(T),
\qquad
\delta P(T)
\simeq
\frac{\pi^2}{6}T^2 g(\mu).
\end{equation}
\begin{equation}
P_\parallel = P_{\rm mat}-\frac{B^2}{2},
\label{eq:Pparallel}
\end{equation}

\begin{equation}
P_\perp = P_{\rm mat}+\frac{B^2}{2}.
\label{eq:Pperp}
\end{equation}
Equations~(82) and~(83) define the no-magnetisation approximation used in this baseline calculation.
The anisotropy included here is the Maxwell-field anisotropy. The matter magnetisation contribution $MB$, which follows from the $B$-dependence of the Landau-quantised matter pressure, is neglected in this baseline.

\begin{figure*}[h]
    \centering
    \includegraphics[width=0.95\textwidth]{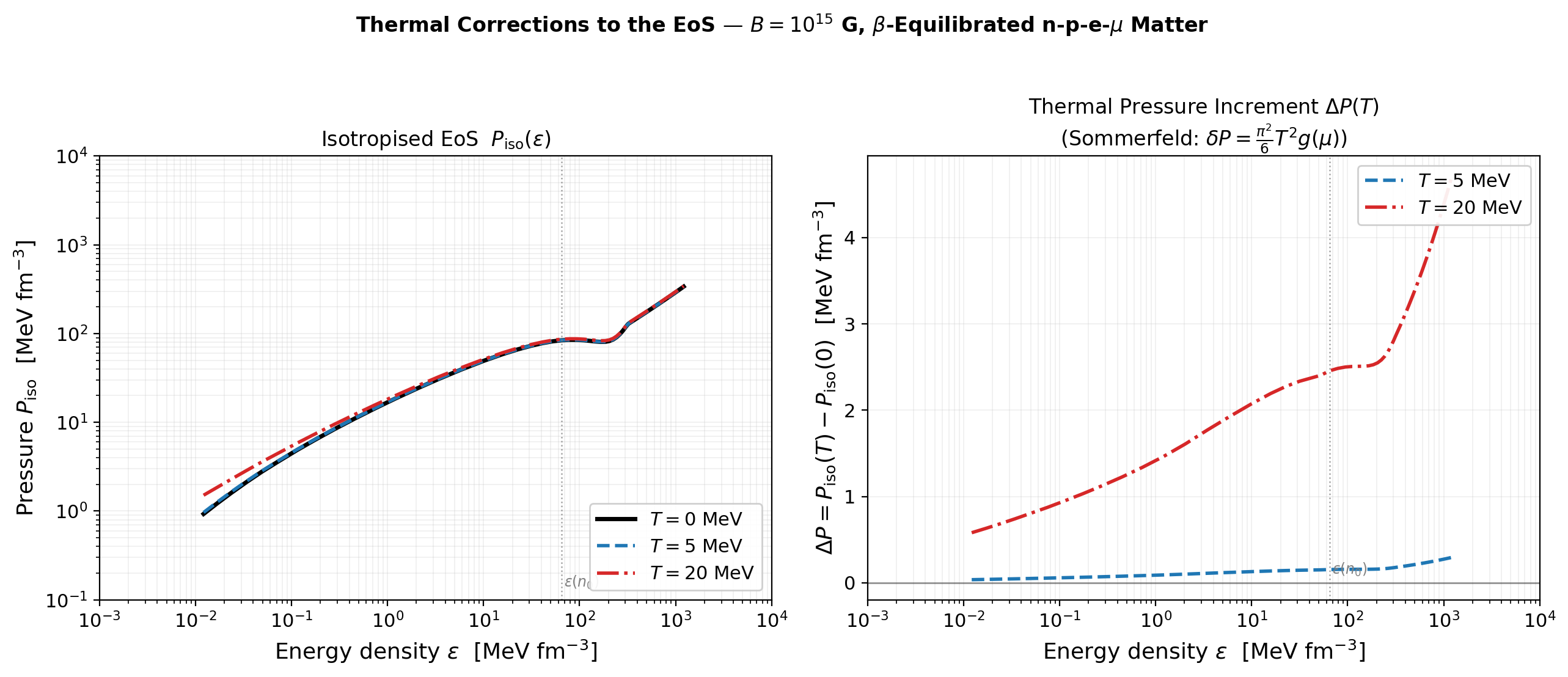}
    \caption{Thermal corrections to the isotropised equation of state
at fixed $B=10^{15}$\,G for $T=0$, $5$, and $20$\,MeV.
\textit{Left panel:} $P_\mathrm{iso}(\varepsilon)$ for the three
temperatures; curves are nearly indistinguishable on the log-log
scale, reflecting the small magnitude of thermal corrections at
neutron-star densities. \textit{Right panel:} Thermal pressure
increment $\Delta P = P_\mathrm{iso}(T)-P_\mathrm{iso}(0)$,
computed via the leading-order Sommerfeld expansion
$\delta P \approx (\pi^2/6)\,T^2\,g(\mu)$, where
$g(\mu)=\partial n/\partial\mu$ is the total density of states
at the Fermi surface evaluated from the self-consistent $T=0$
solution. The expansion is well controlled in the regime
$T \ll \mu$; at saturation density $\mu_n \approx 609$\,MeV,
the thermal correction amounts to approximately $3\%$ of the
total pressure at $T=20$\,MeV. The vertical dotted line marks
the saturation energy density $\varepsilon(n_0)$. The thermal
increment peaks near $\varepsilon(n_0)$ where the density of
states is largest, and decreases at high energy density as
increasing Fermi degeneracy suppresses thermal excitations.}
    \label{fig:thermal_eos}
\end{figure*}

Figure~\ref{fig:thermal_eos} shows the effect of finite
temperature on the EoS at fixed $B = 10^{15}$\,G. Both
panels are computed using the same method: the
zero-temperature self-consistent RMF solution is obtained
first, and the thermal pressure increment is then evaluated
via the leading-order Sommerfeld expansion
$\delta P \approx (\pi^2/6)\,T^2\,g(\mu)$, where
$g(\mu) = \partial n/\partial\mu$ is the total density of
states at the Fermi surface~\cite{prakash1997composition}. This
approximation is valid when $T \ll \mu$; at saturation
density $\mu_n \approx 609$\,MeV, so even at $T = 20$\,MeV
the ratio $T/\mu_n \approx 0.03$ confirms that the
expansion is well controlled. The left panel shows
$P_\mathrm{iso}(\varepsilon)$ at $T = 0$, $5$, and
$20$\,MeV; the curves are nearly indistinguishable on the
logarithmic scale, reflecting the small magnitude of
thermal corrections at neutron-star core densities. The
right panel isolates the increment $\Delta P =
P_\mathrm{iso}(T) - P_\mathrm{iso}(0)$ directly, making
the correction visible. It peaks near $\varepsilon(n_0)$
where the density of states is largest, and decreases at
higher density as Fermi degeneracy suppresses thermal
excitations. At $T = 20$\,MeV and $n_0$, $\Delta P \approx
2.5\,\mathrm{MeV\,fm}^{-3}$, corresponding to
approximately $3\%$ of the zero-temperature pressure ---
physically small but not negligible for proto-neutron star
applications where $T$ can reach $30$--$50$\,MeV in the
first seconds after core bounce~\cite{prakash1997composition}. The
$T^2$ scaling predicted by the Sommerfeld expansion is
confirmed numerically: raising the temperature from $5$ to
$20$\,MeV increases $\Delta P$ by a factor of
$16 = (20/5)^2$, as expected. A full finite-temperature
treatment using the complete Fermi--Dirac integrals
introduced in Sec.~\ref{sec:finite_T} would be required
for $T \gtrsim 30$\,MeV, where the Sommerfeld expansion
begins to break down.

To quantify the degree of pressure anisotropy, we also monitor
\begin{equation}
\Delta
=
\frac{P_{\perp}-P_{\parallel}}{P_{\parallel}+2P_{\perp}}.
\end{equation}
Although magnetisation can in principle be computed from
\begin{equation}
M(B)=\left(\frac{\partial P_{\mathrm{mat}}}{\partial B}\right)_{\{\mu_i\},T},
\end{equation}
it is neglected in the present baseline calculation.

\begin{figure}[H]
    \centering
    \includegraphics[width=0.45\textwidth]{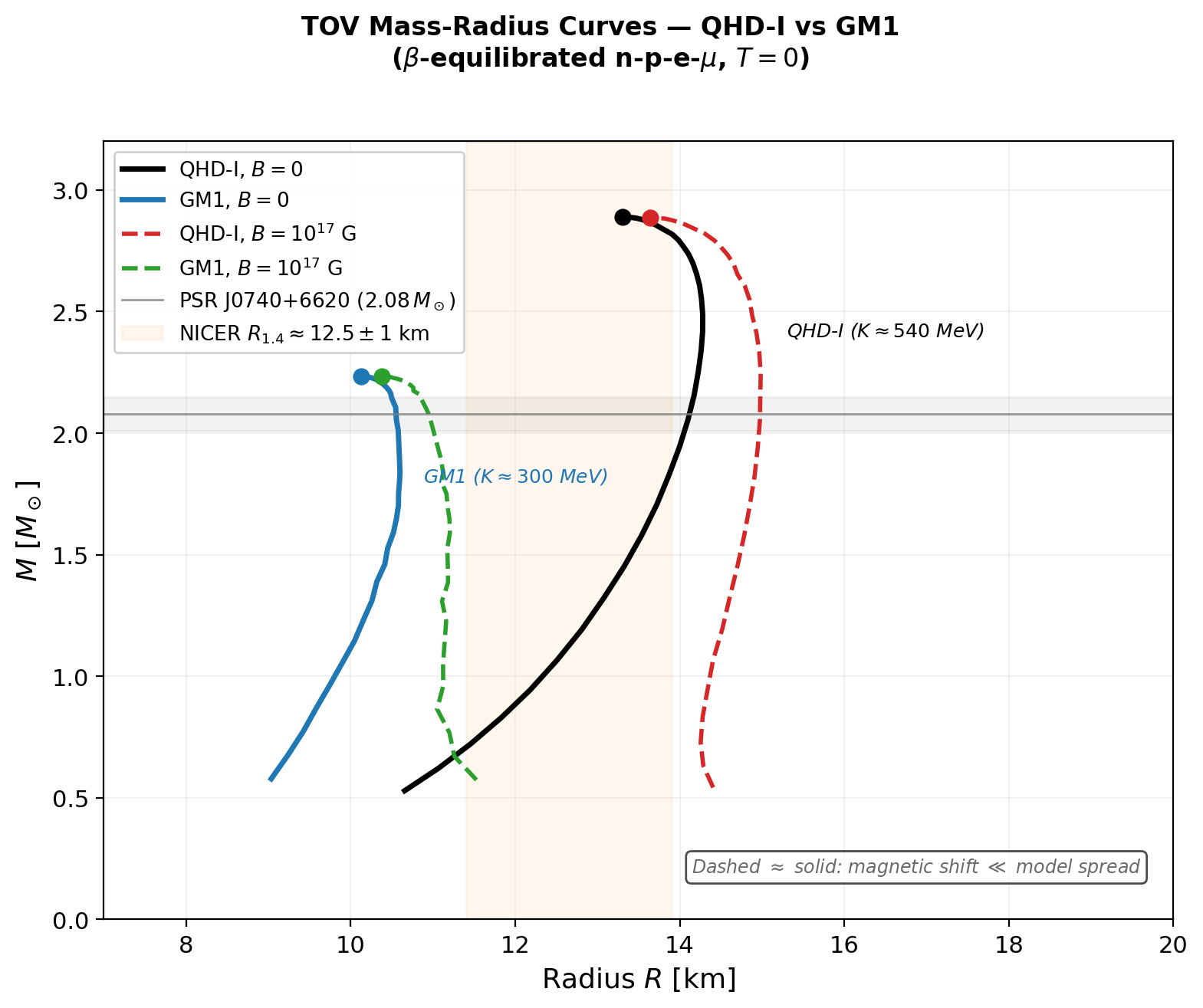}
    \caption{
Pressure anisotropy in magnetised \(\beta\)-equilibrated \(n\)-\(p\)-\(e\)-\(\mu\) matter at \(T=0\), for
\(B=10^{15}\,\mathrm{G}\), \(B=10^{16}\,\mathrm{G}\), and \(B=10^{17}\,\mathrm{G}\).
Left panel: absolute pressure offsets associated with the longitudinal and transverse directions.
Relative to the matter pressure, the Maxwell contribution shifts the longitudinal pressure by
$
P_\parallel-P_{\rm mat}=-\frac{B^2}{2},
$
and the transverse pressure by
$
P_\perp-P_{\rm mat}=+\frac{B^2}{2}.
$
Relative to the isotropised pressure, the corresponding offsets are
$
P_{\rm iso}-P_\parallel=\frac{2B^2}{3},
\qquad
P_{\rm iso}-P_\perp=-\frac{B^2}{3}.
$
Right panel: dimensionless anisotropy parameter
$
\Delta=\frac{P_\perp-P_\parallel}{P_\parallel+2P_\perp}.
$
At low energy density the Maxwell term can dominate over the matter pressure, while at high energy density the matter contribution overwhelms the field term and \(\Delta\to0\). The anisotropy is negligible for \(B\lesssim10^{15}\,\mathrm{G}\) at densities relevant to neutron-star cores, within the no-magnetisation approximation used in this baseline calculation.
}
    \label{fig:anisotropy}
\end{figure}

Figure~\ref{fig:anisotropy} quantifies the pressure anisotropy
directly. Relative to the matter pressure, the Maxwell contribution shifts the longitudinal and transverse pressures by
\begin{equation}
P_\parallel-P_{\rm mat}=-\frac{B^2}{2},
\qquad
P_\perp-P_{\rm mat}=+\frac{B^2}{2}.
\end{equation}
If the splitting is instead measured relative to the isotropised pressure, one obtains
\begin{equation}
P_{\rm iso}-P_\parallel=\frac{2B^2}{3},
\qquad
P_{\rm iso}-P_\perp=-\frac{B^2}{3}.
\end{equation} 

These splittings are independent of density in the no-magnetisation approximation and scale as \(B^2\). Thus, increasing the field from \(10^{15}\,\mathrm{G}\) to \(10^{17}\,\mathrm{G}\) increases the Maxwell contribution by four orders of magnitude. At \(B=10^{17}\,\mathrm{G}\), the Maxwell-induced pressure splitting remains small compared with the characteristic energy density near saturation, but it can become non-negligible compared with the matter pressure at sufficiently low densities. The right panel shows that the dimensionless anisotropy parameter \(\Delta\) peaks at low energy density and declines steeply toward zero above nuclear saturation density.

\section{Discussion and Conclusion}

We have developed and implemented a self-consistent relativistic mean-field framework for magnetised neutron-star matter in $\beta$ equilibrium, incorporating Landau quantisation of all charged species and a complete $n$-$p$-$e$-$\mu$ composition. The framework reproduces the \(B=0\) saturation properties of the QHD-I parameterisation and provides a self-consistent \(T=0\) RMF baseline equation of state over a range of baryon densities and prescribed magnetic-field strengths. Magnetic pressure anisotropy is included through the Maxwell contribution in the no-magnetisation approximation, while finite-temperature effects are treated perturbatively in the degenerate regime.

The main physical conclusions are as follows. First, for magnetic fields of order $B=10^{15}\,\mathrm{G}$, the Landau-level spacing remains negligible compared with the Fermi energy, and the equation of state is effectively indistinguishable from the unmagnetised limit. Visible departures emerge only for $B\gtrsim10^{17}\,\mathrm{G}$, where the charged sector begins to enter the strongly quantising regime. Second, the scalar density remains systematically suppressed relative to the vector density, reflecting the relativistic factor $M^*/E^*$ and confirming the expected many-body structure of RMF matter. Third, pressure anisotropy, quantified by
\begin{equation}
\Delta=\frac{P_\perp-P_\parallel}{P_\parallel+2P_\perp},
\end{equation}
grows with magnetic-field strength at low energy density but becomes negligible at neutron-star core densities, where the matter pressure dominates over the Maxwell term. Fourth, thermal corrections at temperatures relevant to proto-neutron stars ($T\lesssim20\,\mathrm{MeV}$) remain modest, amounting to approximately $3\%$ of the total pressure near saturation density and following the expected Sommerfeld scaling
\begin{equation}
\delta P \approx \frac{\pi^2}{6}T^2 g(\mu),
\end{equation}
which confirms that the degenerate approximation is well controlled in this regime.

As an exploratory application, we used the isotropised equation of state to estimate the sensitivity of neutron-star mass--radius relations to magnetic fields. Within the isotropised-TOV proxy, the inferred changes in the maximum mass and canonical radius remain very small up to \(B\lesssim 10^{17}\,\mathrm{G}\). These results should be interpreted only as exploratory sensitivity estimates of the baseline EoS, not as a substitute for a fully anisotropic magnetised-star calculation.

The present framework has several deliberate limitations that define the scope of future work. Anomalous magnetic moments have been neglected, although they are expected to become quantitatively important for $B\gtrsim10^{18}\,\mathrm{G}$. The magnetic field is treated as an
externally prescribed parameter rather than a dynamical
quantity. Realistic magnetic-field profiles require coupling
the microscopic EoS to self-consistent stellar magnetic-field
configurations~\cite{dexheimer2017magnetic,glampedakis2016freedom}.
In addition, explicit coupling to ambipolar diffusion,
Hall drift, and Ohmic dissipation remains to be developed. In the present study, the QHD-I parameterisation
is intentionally retained as a stiff benchmark baseline,
while the GM1 parameterisation provides a more realistic
comparison with reduced incompressibility. More comprehensive
RMF treatments may include hyperonic degrees of freedom
and compare several parameterisations such as GM1, NL3,
TM1, FSUGold, and IU-FSU~\cite{miyatsu2013equation, miyatsu2013new}.
Even so, neither model includes hyperons or quark degrees of freedom,
both of which are expected to soften the equation
of state at densities above \(2\text{--}3n_0\). Extending the present
framework to nonlinear or density-dependent RMF
parameterisations, and ultimately to more realistic
multi-component compositions, is therefore a necessary
next step. Future extensions should also compare the
present baseline with modern calibrated RMF interactions
such as FSUGold~\cite{toddrutel2005neutron}.

The present work should thus be viewed as a minimal, internally consistent baseline for magnetised dense matter: sufficiently complete to isolate the leading effects of Landau quantisation, pressure anisotropy, and moderate thermal corrections, yet simple enough to serve as a benchmark for future extensions. In that sense, it provides a controlled starting point for subsequent studies of neutron-star cooling, stability, and fully anisotropic stellar structure in strong magnetic fields.
\newpage
\bibliography{main.bib}

@article{duncan1992formation,
  title={Formation of very strongly magnetized neutron stars-Implications for gamma-ray bursts},
  author={Duncan, Robert C and Thompson, Christopher},
  journal={Astrophysical Journal, Part 2-Letters (ISSN 0004-637X), vol. 392, no. 1, June 10, 1992, p. L9-L13. Research supported by NSERC.},
  volume={392},
  pages={L9--L13},
  year={1992}
}

@article{turolla2015magnetars,
  title={Magnetars: the physics behind observations. A review},
  author={Turolla, Roberto and Zane, Silvia and Watts, AL},
  journal={Reports on Progress in Physics},
  volume={78},
  number={11},
  pages={116901},
  year={2015},
  publisher={IOP Publishing}
}

@book{glendenning2012compact,
  title={Compact stars: Nuclear physics, particle physics and general relativity},
  author={Glendenning, Norman K},
  year={2012},
  publisher={Springer Science \& Business Media}
}

@article{serot1986advance,
  title={Advance in nuclear physics},
  author={Serot, Brian D},
  journal={The Relativistic Nuclear Many-Body Problem},
  volume={16},
  year={1986},
  publisher={Plenum Press}
}

@article{potekhin2015neutron,
  title={Neutron stars—cooling and transport},
  author={Potekhin, Alexander Y and Pons, Jos{\'e} A and Page, Dany},
  journal={Space Science Reviews},
  volume={191},
  pages={239--291},
  year={2015},
  publisher={Springer}
}

@article{harding2006physics,
  title={Physics of strongly magnetized neutron stars},
  author={Harding, Alice K and Lai, Dong},
  journal={Reports on Progress in Physics},
  volume={69},
  number={9},
  pages={2631},
  year={2006},
  publisher={IOP Publishing}
}

@article{walecka1974theory,
  title={A theory of highly condensed matter},
  author={Walecka, JD},
  journal={Annals of Physics},
  volume={83},
  number={2},
  pages={491--529},
  year={1974},
  publisher={Elsevier}
}

@article{glendenning1991reconciliation,
  title={Reconciliation of neutron-star masses and binding of the $\Lambda$ in hypernuclei},
  author={Glendenning, NK and Moszkowski, SA},
  journal={Physical review letters},
  volume={67},
  number={18},
  pages={2414},
  year={1991},
  publisher={APS}
}

@article{rabhi2011warm,
  title={Warm and Dense Stellar Matter under Strong Magnetic Fields},
  author={Rabhi, Aziz and Panda, P. K. and Provid{\^e}ncia, Constan{\c{c}}a},
  journal={Physical Review C},
  volume={84},
  number={3},
  pages={035803},
  year={2011},
  doi={10.1103/PhysRevC.84.035803}
}

@article{miyatsu2013new,
  title={A new equation of state for neutron star matter with nuclei in the crust and hyperons in the core},
  author={Miyatsu, Tsuyoshi and Yamamuro, Sachiko and Nakazato, Ken'ichiro},
  journal={The Astrophysical Journal},
  volume={777},
  number={1},
  pages={4},
  year={2013},
  publisher={The American Astronomical Society}
}

@article{miyatsu2013equation,
  title={Equation of state for neutron stars in SU (3) flavor symmetry},
  author={Miyatsu, Tsuyoshi and Cheoun, Myung-Ki and Saito, Koichi},
  journal={Physical Review C—Nuclear Physics},
  volume={88},
  number={1},
  pages={015802},
  year={2013},
  publisher={APS}
}

@article{toddrutel2005neutron,
  title={Neutron-Rich Nuclei and Neutron Stars: A New Accurately Calibrated Interaction for the Study of Neutron-Rich Matter},
  author={Todd-Rutel, B. G. and Piekarewicz, J.},
  journal={Physical Review Letters},
  volume={95},
  number={12},
  pages={122501},
  year={2005},
  doi={10.1103/PhysRevLett.95.122501}
}

@article{dexheimer2017magnetic,
  title={The Magnetic Field Distribution in Strongly Magnetized Neutron Stars},
  author={Dexheimer, V. and Franzon, B. and Gomes, R. O. and Farias, R. L. S. and Avancini, S. S. and Schramm, S.},
  journal={Physics Letters B},
  volume={773},
  pages={487--491},
  year={2017},
  doi={10.1016/j.physletb.2017.09.008}
}

@article{rabhi2008stellar,
  title={Stellar Matter with a Strong Magnetic Field within Density-Dependent Relativistic Models},
  author={Rabhi, Aziz and Provid{\^e}ncia, Constan{\c{c}}a and Provid{\^e}ncia, Jo{\~a}o da},
  journal={Journal of Physics G: Nuclear and Particle Physics},
  volume={35},
  number={12},
  pages={125201},
  year={2008},
  doi={10.1088/0954-3899/35/12/125201}
}

@article{dong2013magnetization,
  title={Magnetization of Neutron Star Matter},
  author={Dong, Jianmin and Zuo, Wei and Gu, Jianzhong},
  journal={Physical Review D},
  volume={87},
  number={10},
  pages={103010},
  year={2013},
  doi={10.1103/PhysRevD.87.103010}
}

@article{ferrer2010equation,
  title={Equation of State of a Dense and Magnetized Fermion System},
  author={Ferrer, Efrain J. and de la Incera, Vivian and Keith, Jason P. and Portillo, Israel and Springsteen, Paul L.},
  journal={Physical Review C},
  volume={82},
  number={6},
  pages={065802},
  year={2010},
  doi={10.1103/PhysRevC.82.065802}
}

@article{glampedakis2016freedom,
  title={The Freedom to Choose Neutron Star Magnetic Field Equilibria},
  author={Glampedakis, Kostas and Lasky, Paul D.},
  journal={Monthly Notices of the Royal Astronomical Society},
  volume={463},
  number={3},
  pages={2542--2552},
  year={2016},
  doi={10.1093/mnras/stw2128}
}

@article{broderick2000equation,
  title={The Equation of State of Neutron-Star Matter in Strong Magnetic Fields},
  author={Broderick, Avery and Prakash, Madappa and Lattimer, James M.},
  journal={The Astrophysical Journal},
  volume={537},
  number={1},
  pages={351--367},
  year={2000},
  doi={10.1086/309010}
}

@article{strickland2012bulk,
  title={Bulk properties of a Fermi gas in a magnetic field},
  author={Strickland, Michael and Dexheimer, Veronica and Menezes, Debora P},
  journal={Physical Review D—Particles, Fields, Gravitation, and Cosmology},
  volume={86},
  number={12},
  pages={125032},
  year={2012},
  publisher={APS}
}

@article{miller2021radius,
  title={The radius of PSR J0740+ 6620 from NICER and XMM-Newton data},
  author={Miller, M Coleman and Lamb, FK and Dittmann, AJ and Bogdanov, S and Arzoumanian, Z and Gendreau, KC and Guillot, S and Ho, WCG and Lattimer, JM and Loewenstein, M and others},
  journal={The Astrophysical Journal Letters},
  volume={918},
  number={2},
  pages={L28},
  year={2021},
  publisher={The American Astronomical Society}
}

@article{fonseca2021refined,
  title={Refined mass and geometric measurements of the high-mass PSR J0740+ 6620},
  author={Fonseca, Emmanuel and Cromartie, H Thankful and Pennucci, Timothy T and Ray, Paul S and Kirichenko, A Yu and Ransom, Scott M and Demorest, Paul B and Stairs, Ingrid H and Arzoumanian, Zaven and Guillemot, Lucas and others},
  journal={The Astrophysical Journal Letters},
  volume={915},
  number={1},
  pages={L12},
  year={2021},
  publisher={The American Astronomical Society}
}

@article{prakash1997composition,
  title={Composition and structure of protoneutron stars},
  author={Prakash, Madappa and Bombaci, Ignazio and Prakash, Manju and Ellis, Paul J and Lattimer, James M and Knorren, Roland},
  journal={Physics Reports},
  volume={280},
  number={1},
  pages={1--77},
  year={1997},
  publisher={Elsevier}
}
\bibliographystyle{abbrv}

\end{document}